\documentclass[12pt]{iopart}

\usepackage[utf8]{inputenc}
\usepackage[T1]{fontenc}
\usepackage[english]{babel}
\usepackage[dvipsnames]{xcolor}
\usepackage[pdftex]{graphicx}
\usepackage{amsthm,amsfonts,amssymb,amsopn,amsmath}
\usepackage{dsfont}
\usepackage{layouts,enumerate}
\usepackage{blkarray}
\usepackage{mathtools}
\usepackage{url,hyperref,cite}
\usepackage{pgf,tikz,hhline}
\usepackage{iopams}
\usepackage{calc,xstring,ifthen}
\usepackage{pdftexcmds,adjustbox}
\usetikzlibrary{external,arrows,calc,matrix,backgrounds,fit,positioning,
	decorations.pathmorphing,decorations.markings}
\usepackage{bm}
\usepackage{cleveref}
\usepackage{caption,float}

\def\bra#1{\mathinner{\langle{#1}|}}
\def\ket#1{\mathinner{|{#1}\rangle}}

\def\ul#1{\underline{#1}}

\def\mbb#1{\mathbb{#1}}

\def\one{\mathds{1}}
\def\mA{\mathfrak{A}}
\def\mB{\mathfrak{B}}

\newcommand{\ZZ}{\mathbb{Z}}
\newcommand{\NN}{\mathbb{N}}

\newcommand{\CC}{\mathbb{C}}

\usepackage[parfill]{parskip}
\begin{document}

\title{Operator spreading in quantum hardcore gases}

\date{\today}

\author{Marko Medenjak$^{1}$}
\address{$^1$ Department of Theoretical Physics, University of Geneva,
	24 quai Ernest-Ansermet, 1211 Geneva, Switzerland}
\ead{Marko.Medenjak@unige.ch}

\begin{abstract} 
 In this article we study a set of integrable quantum cellular automata{{, the quantum hardcore gases (QHCG)}}, with an arbitrary local Hilbert space dimension, and discuss the matrix product ansatz based approach for solving the dynamics of local operators analytically. Subsequently, we focus on the dynamics of operator spreading, in particular on the out-of-time ordered correlation functions (OTOCs),  operator weight spreading and {{operators space entanglement entropy (OSEE)}}. All of the quantities were conjectured to provide signifying features of integrable systems and quantum chaos. We show that in  QHCG OTOCs spread diffusively and that in the limit of the large local Hilbert space dimension  they increase  linearly with time, despite their integrability. On the other hand, it was recently conjectured that operator weight front, which is associated with the extent of operators, spreads diffusively in both, integrable and generic systems, but its decay seems to differ in these two cases \cite{lopez2021operator}.  We observe that the spreading of the operator weight front in  QHCG is markedly different from chaotic, generic integrable and free systems, as the front freezes in the long time limit. Finally, we discuss the OSEE in QHCG and show that it grows at most logarithmically with time in accordance with the conjectured behaviour for interacting integrable systems \cite{Alba2019Operator}.
\end{abstract}
\maketitle
\section{{Introduction}}

In recent decades there were numerous attempts to distinguish between ergodic, solvable interacting and free dynamics. One of the simplest ways to do it is to study conservation laws and their structure \cite{IMPZ}. Nevertheless, it is not clear how the existence of these conservation laws impacts the complexity of the time evolution of local observables. In this article we will focus on three objects which quantify how the operators spread in quantum many-body system. These quantities are of particular interest since they were conjectured to distinguish between integrable and chaotic dynamics.

One of the salient features of chaos in classical systems is the sensibility of the dynamics to initial conditions, which is characterized by the exponential separation of the close-by phase space trajectories. This separation is characterized by Lyapunov exponents. In integrable systems, on the other hand, Lyapunov exponents vanish. This means that in order to capture the dynamics of chaotic systems up to time $t$ exactly, exponential resources in $t$ are required, while this is not the case in integrable systems. In the last decade much of the effort has been devoted to the study of many-body quantum chaos, following the work of \cite{maldacena2016bound}, where the authors proposed the bound on quantum chaos by considering the quantum analog of classical trajectory separation, the Out-of-time ordered correlation functions (OTOCs) \cite{larkin1969quasiclassical,khemani2018velocity,gopalakrishnan2018hydrodynamics,nahum2018operator,von2018operator,Chan2018Solution,gopalakrishnan2018operator,pappalardi2018scrambling,foini2019eigenstate,bernard2021dynamics,bensa2021two,lerose2020bridging}. However, it can be shown that if the interaction is local, and the number of local states finite OTOCs of local observables cannot exhibit exponential growth \cite{kukuljan2017weak,von2018operator}. Therefore in such cases OTOCs are not able to distinguish between integrability and chaos. 

In this article we address the question whether such distinction can be provided by OTOCs in the limit of infinitely many local degrees of freedom, by computing them analytically in the set of integrable systems with a varying local Hilbert space dimension. We show that in quantum hardcore gases (QHCG) OTOCs increase linearly with time in the limit of a large local Hilbert space dimension.

Operator front, which is the second quantity that we consider, quantifies the spreading of the distribution of the right-most/left-most part of the local operators. In random unitary circuits operator front was related to OTOCs \cite{von2018operator}, but provides  distinct information otherwise. This prompted its study as a distinguishing feature between free, interacting integrable, and chaotic dynamics \cite{lopez2021operator}.  In this article we obtain analytical expressions for the spreading of operators in the set of aforementioned quantum integrable automata, and show that its dynamics does not agree with the conjecture for interacting integrable systems provided in \cite{lopez2021operator}.

Finally, we focus on the operator space entanglement entropy (OSEE), which quantifies the complexity of the matrix product based techniques for simulating the dynamics of local observables. Namely, linear increase of the operator entanglement implies that exponential in time resources are required in order to exactly capture the dynamics of local observables using the matrix product ansatz based methods. Operator entanglement is expected to grow linearly in generic systems, while it saturates in free systems \cite{zanardi2001entanglement,pizorn2009operator,Alba2019Operator}. In \cite{Alba2019Operator} it was conjectured that in interacting integrable systems the increase of OSEE is logarithmic, which implies algebraic increase of complexity with time. In the present paper we calculate an upper bound on OSEE  showing that the increase is maximally logarithmic in time in support of the aforementioned conjecture.

In this work we focus on the dynamics of quantum cellular automata.
Cellular automata are many-body systems where both time and space are discrete and time evolution deterministic. Recently they attracted a significant amount of attention in the study of strongly interacting out-of-equilibrium systems \cite{bobenko1993two,PhysRevLett.119.110603,klobas2019time,1742-5468-2018-12-123202,10.21468/SciPostPhys.6.6.074,friedman2019integrable,10.21468/SciPostPhysCore.2.2.010,Klobas_2020,klobas2021exact,Alba2019Operator,PhysRevLett.126.160602,PhysRevE.102.062107,bertini2021entanglement,buvca2021rule,balazs20212,balazs2021,gombor2021superintegrable}, as they admit analytical solutions of the dynamics, which goes well beyond what is typically possible even in integrable quantum systems.  There are two notable examples of such systems, the Rule 54 chain \cite{klobas2019time}, and the two species hardcore gas \cite{10.21468/SciPostPhys.6.6.074}, which admit closed form solutions of the dynamics for local operators in the matrix product ansatz form. In this paper we consider a set of integrable quantum cellular automata, which can be understood as the quantum multi-particle generalizations of the model introduced in \cite{10.21468/SciPostPhys.6.6.074}. They allow us to rigorously discern the dynamics of three quantities associated with the complexity of operators, which were previously conjectured to distinguish between chaotic and integrable dynamics: OTOCs, operator front spreading, and OSEE.

\section{{Outline of the paper and main results}}
In section \ref{QHCG} we discuss the notion of quantum cellular automata, and  their relevance as paradigmatic interacting models with exactly solvable dynamics. We introduce a new set of models with an arbitrary number of particle species and demonstrate their integrability \eqref{int}.

In section \ref{tMPA} we introduce an algebraic approach for solving the dynamics of the cellular automata by the matrix product ansatz form of local observables, which comprises the bulk, the boundary and initial algebraic relations \ref{cond}. We demonstrate on a simple example how the approach can be used to discern whether or not a given operator density appears in the time evolution of the local operator, without knowing an explicit solution of aforementioned algebraic conditions. In subsequent sections analogous approach is employed to discern the dynamics of more complicated objects.

 The first quantities that we focus on is how the OTOCs spread in QHCGs. In section \ref{OTOC} we calculate the spreading of OTOCs \eqref{OTOC_f}. We show that they spread diffusively and that in the limit of the infinite local Hilbert space dimension they increase linearly with time.

In section \ref{OW} we obtain an exact expression for the operator weight spreading in quantum hard-core gases \eqref{R1}, \eqref{R2}. We show that, contrary to expectations for interacting integrable models \cite{lopez2021operator}, the operator weight in QHCG moves with a constant velocity but remains localized in time.

Finally, in section \ref{opent} we discuss the operator entanglement spreading in QHCG and provide a strict upper bound that shows that entanglement increases at-most logarithmically with time, as predicted in \cite{Alba2019Operator}.

In table \ref{sumup} we summarize the main results of the paper on the dynamics of operators in QHCG and show how they compare to the conjectures for interacting integrable systems in the literature \cite{lopez2021operator,Alba2019Operator}. In particular we focus on the late time dynamics of the three aforementioned quantities: the increase of OTOCs (depending on the local Hilbert space dimension $q$), the decay of the operator front, and the increase of the OSEE.
\newline

\begin{center}
	\label{sumup}
	\begin{tabular}{ |c|c|c|c|c|c| } 
		\hline
		 & Free & Chaotic & Integrable & QHCG \\
		\hline
		OTOC& Const. & $\exp( t)$ & Const. & Const. ($q<\inf$), $t$ ($q\to\infty$) \\ 
		Operator front &  $t^{-2/3}$ & $t^{-1/2}$ & $t^{-3/4}$ & Const.\\ 
		OSEE  & Const. & $t$ & $\log(t)$ & $\log(t)$ \\ 
		\hline
	\end{tabular}
\end{center}

\section{Cellular automata and the quantum hardcore gas\label{QHCG}}
\begin{figure}
	\centering
	\includegraphics[width=0.5\textwidth]{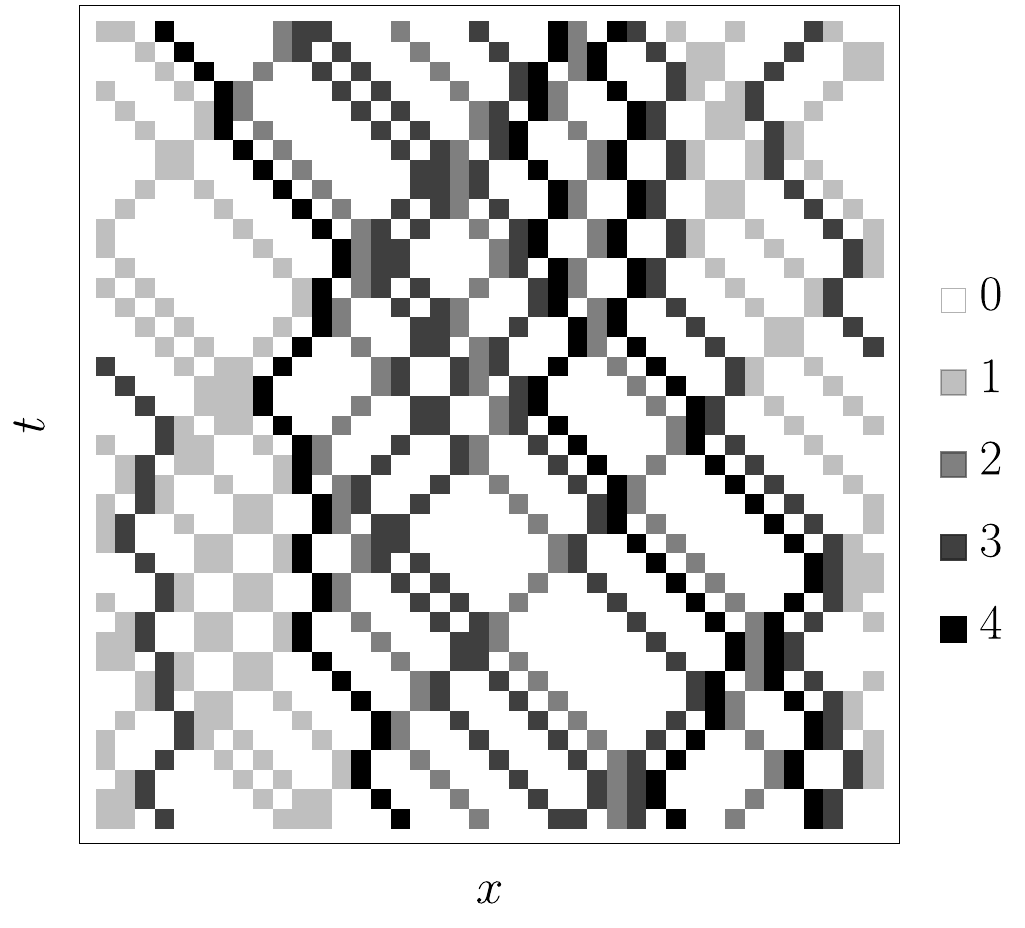}
	\caption{\label{fig:dyn} The dynamics of   QHCG with $4$ particle species. Vacancies correspond to the number $0$.
	}
\end{figure}

{ Reversible cellular automata are local deterministic systems that are discrete in space and time, where every configuration $\underline{s}=\{\cdots,s_{-1},s_0,s_1,\cdots\}$ is mapped to a distinct configuration $\underline{s}'=\{\cdots,s_{-1}',s_0',s_1',\cdots\}$ and vice verse, and the updated state $s_i'$ at site $i$ depends only on nearby points at the previous time step. If we interpret the configuration $\underline{s}$ as a pointer state in the Hilbert space $\ket{\underline{s}}=\ket{\cdots s_{-1} s_0 s_1,\cdots}$ the cellular automaton dynamics induces quantum time evolution, which corresponds to a linear map that maps the configuration $\ket{\underline{s}}=\ket{\cdots s_{-1} s_0 s_1\cdots}$ to $\ket{\underline{s}'}=\ket{\cdots s_{-1}'s_0's_1'\cdots}$. Unitarity of the quantum evolution is due to the reversibility and the bijective nature of the process. Note that such dynamics is rather restrictive as it preserves the subspace of diagonal observables, which are in general of the form $\sum \lambda_{\ul{s}}\ket{\underline{s}}\bra{\underline{s}}$. This is simply a consequence of both $\ket{\underline{s}}$ and $\bra{\underline{s}}$ being mapped to corresponding pointer states $\ket{\underline{s}'}$ and $\bra{\underline{s}'}$. Nevertheless, one can consider the induced dynamics of off-diagonal observables, which are inherently quantum. Despite this simplification such systems allow us to study fundamental features of quantum many-body systems, such as equilibration rigorously \cite{PhysRevLett.126.160602}. 

In the present article we focus on cellular automata in which the dynamics is governed by the brick-wall structure, see Figure~\ref{bvs2}.
\begin{figure}
	\centering
	\includegraphics[width=0.8\textwidth]{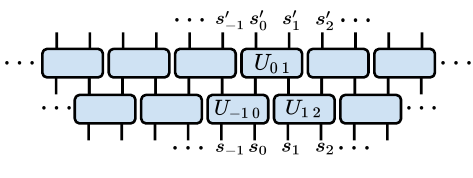}
	\caption{\label{bvs2} The dynamics of QHCG is induced by the staggered local deterministic unitary gates $U_{i i+1}$, acting on neighboring bonds.
	}
\end{figure}
The local gate
\begin{equation}
	\label{lprop}
	U_{12} \ket{s s'}=(1-\delta_{s,0})(1-\delta_{s',0})\ket{s s'}+(\delta_{s,0}+\delta_{s',0}-\delta_{s,0}\delta_{s',0})\ket{s' s},
\end{equation}
corresponds to the hard-core interaction between the $(q-1)$ different particle types $\{\ket{1},\dots,\ket{q-1}\}$ and freely propagating vacancies $\ket{0}$, see Fig~\ref{fig:dyn}. The local gate can be understood as a permutation between vacancies $A=\{0\}$ and the set of particles $B=\{1,2,\cdots q-1\}$
\begin{equation}
	\label{lpr}
U=\begin{blockarray}{ccccc}
   	& AA & AB & BA & BB  \\
	\begin{block}{c(cccc)}
	AA	& 1 & 0 & 0 & 0  \\
	AB &	0 & 0 & \one & 0 \\
	BA &	0 & \one & 0 & 0  \\
	BB &	0 & 0 & 0 & \one  \\
	\end{block}
\end{blockarray}
\end{equation}

The propagator is obtained by stacking two layers of local gates on top of each-other as in Figure~\ref{bvs2}
\begin{equation}
	\label{prop}
	\mathcal{U}=\prod_{i\in 2\ZZ} U_{ii+1}\prod_{i\in 2\ZZ+1} U_{ii+1}.
\end{equation}
Integrability of the process follows, in general, from the commutativity of the propagator $\mathcal{U}$ with the family of transfer matrices
\begin{equation}
	[\mathcal{U},T(\lambda)]=[T(\mu),T(\lambda)]=0,
\end{equation}
which typically allow us to construct local conservation laws \cite{IMPZ}.
Following the derivation in \cite{10.21468/SciPostPhys.6.6.074} we can show that the $R$-matrix defined as 
\begin{equation}
	R_{12}(\lambda)=P_{12}\frac{\one-\lambda U_{12}}{1+\lambda},
\end{equation}
where $P_{12}$ is the permutation and $\one$ the identity operator, can be used to construct the set of commuting transfer matrices $T(\lambda)$
\begin{equation}
	T(\lambda)=\text{tr}(R_{01}(\lambda+\eta/2)R_{02}(\lambda-\eta/2)\cdots),
\end{equation}
which depend on the spectral parameter $\lambda$ and staggering $\eta$.
Commutativity of transfer matrices for different values of the spectral parameter $\lambda$ arises as a consequence of the braid relations for the local propagator $U$
\begin{equation}
	U_{12}U_{23}U_{12}=U_{23}U_{12}U_{23},
\end{equation}
which follows immediately from the permutation like structure \eqref{lpr}.
The dynamics of the quantum cellular automaton can then be obtained by considering the large staggering limit of the two transfer matrices
\begin{equation}
	\label{int}
		\mathcal{U}=\lim_{\eta\to\infty}T(-\eta/2)^\dagger T(\eta/2).
\end{equation} }
An example of the time evolution of QCHG for a random initial configuration with $4$ particle species is shown in Figure~\ref{fig:dyn}. For further discussion of integrability of related cellular automata  we refer the reader to the recent work \cite{gombor2021superintegrable}.
\section{{tMPA for the local unitary circuits} \label{tMPA}}
Matrix product ansatz is in general used to encode correlations in states, density matrices, or operators in the auxiliary space. Locality of the ansatz allows us to solve the local quantum dynamics fully, by finding the solution to the set of bulk, boundary and initial algebraic relations, generalizing the classical approach introduced in \cite{10.21468/SciPostPhys.6.6.074}. In this section we will first introduce the set of conditions and show how they lead to the solution of dynamics. Afterwards we will discuss how the algebraic relations themselves can be used to calculate physical observables analytically.
Note that in general the finite time dynamics of local observables in integrable models cannot be discerned analytically and there are only few examples where such solutions exist  \cite{klobas2019time,10.21468/SciPostPhys.6.6.074}.

\begin{figure}
	\centering
	\includegraphics[width=0.5\textwidth]{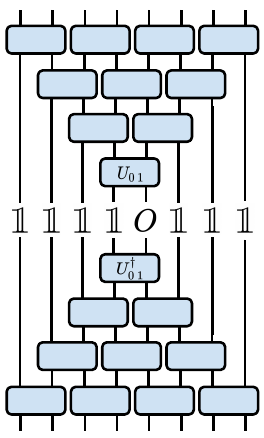}
	\caption{\label{loc_op} Time evolution of the operator is reduced to the light-cone since the gates $U_{i\,i+1}$ and $U^\dagger_{i\,i+1}$ outside of the light-cone compensate. Operator at time $t$ can therefore be obtained by applying the set of gates inside of the light-cone at time $t-1$ to the operator at time $t-1$, as indicated by equation \eqref{disc_dyn}.
	}
\end{figure}

Any operator $O$ which is at time $t=0$ localized at site $1$ can be expressed in terms of auxiliary space matrices $A_{ij}(t)$ and $B_{ij}(t)$, and left and right auxiliary space boundary vectors $\bra{L_{ij}(t)},\ \ket{R(t)}$ restricted to the light-cone
\begin{equation}
	\label{ansatz}
	O(t)=\sum_{\underline{s},\underline{s}'}\bra{L_{s_{-t+1}s'_{-t+1}}(t)}  A_{s_{-t+2}s'_{-t+2}}(t) \cdots B_{s_{t-1}s'_{t-1}}(t) A_{s_{t}s'_{t}}(t) \ket{R(t)} E^{s_{-t+1}s'_{-t+1}}_{-t+1}\dots E^{s_t s_t}_{t},
\end{equation} 
where $E^{i j}_x=\cdots\otimes\one\otimes\underbrace{E^{ij}}_{\text{site x}}\otimes\one\otimes\cdots$ for $i,j\in \{0,...,q-1\} $ are local operator basis elements, with $E^{ij}=\ket{i}\bra{j}$,  and the sum runs over all the basis elements within the light-cone. The time evolved operator can be written more compactly in terms of matrices and boundary vectors acting on both physical and auxiliary space
\begin{equation}
	\mathfrak{A}(t)=\sum_{i,j} A_{ij}(t)E^{ij},\quad \mathfrak{B}(t)=\sum_{i,j} B_{ij}(t) E^{ij},\quad \bra{\mathfrak{L}(t)}=\sum_{i,j} \bra{L_{ij}(t)} E^{ij}
\end{equation}
as
\begin{equation}
	O(t)=\cdots \otimes\one\otimes \bra{\mathfrak{L}(t)}\dot{\otimes}\underbrace{\mathfrak{A}(t)\dot{\otimes}\mathfrak{B}(t)\dot{\otimes}\cdots\dot{\otimes}\mathfrak{B}(t)\dot{\otimes} \mathfrak{A}(t)}_{2t-1}\cdot\ket{R(t)}\otimes{\one}\otimes\cdots,
\end{equation}
where $\dot{\otimes}$ corresponds to the tensor product over the physical space and multiplication in the auxiliary space.
Note that the ansatz \eqref{ansatz} is restricted to the light-cone of the width $2t$, which means that the dynamics of the local operator corresponds to
\begin{equation}
	\label{disc_dyn}
	O(t+1)=(U_{-t,-t+1} \cdots U_{t-2,t-1} U_{t,t+1} )\, O(t)\, (U_{t,t+1}^\dagger  U_{t-2,t-1}^\dagger  \cdots U_{-t,-t+1}^\dagger),
\end{equation}
see Figure~\ref{loc_op}.
The main idea is to write algebraic conditions for matrices $\mathfrak{A}(t)$, $\mathfrak{B}(t)$ and  boundary vectors $\bra{\mathfrak{L}(t)}$, $\ket{R(t)}$
such that equation \eqref{disc_dyn} is satisfied at all times. There are three algebraic conditions which ensure that the ansatz \eqref{ansatz} provides the solution to \eqref{disc_dyn}
\begin{enumerate}
	\label{cond}
	\item Bulk permutation condition
	\begin{equation}
		\label{bp_cond}
		U  (\mA(t)\otimes\mB(t)) U^\dagger=\mB(t+1)\otimes\mA(t+1),
	\end{equation}
	\item Boundary conditions
	\begin{align}
		\label{nene}
		\begin{split}
		\bra{\mathfrak{L}(t+1)}_{-t}\,\mA(t+1)_{-t+1}&= U_{-t,-t+1}(\one\otimes  \bra{\mathfrak{L}(t)}_{-t+1})U_{-t,-t+1}^\dagger,\\
	\one \ket{R(t)} &= \mB(t)  \ket{R(t+1)},
\end{split}	
\end{align}
	\item Initial condition
	\begin{eqnarray}
		\label{init}
		O(0)=\bra{\mathfrak{L}(0)}R(0)\rangle
	\end{eqnarray}
\end{enumerate}

\begin{figure}
	\centering
	\includegraphics[width=1\textwidth]{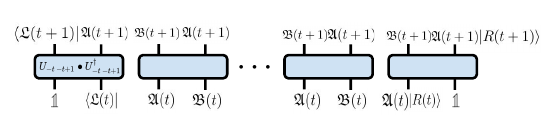}
	\caption{\label{loc_op1} Graphical proof that the bulk, boundary, and initial algebraic conditions provide the solution of the dynamics. The boxes represent both the application of the unitary matrix $U$ and its adjoint $U^\dagger$. If at time $t$ we have the matrix product ansatz given by matrices $\mB(t)$, $\mA(t)$, $\bra{\mathfrak{L}(t)}$ and $\ket{R(t)}$ then we can use the bulk and boundary conditions to show that the matrix ansatz at time $t+1$: $\mB(t+1)$, $\mA(t+1)$, $\bra{\mathfrak{L}(t+1)}$ and $\ket{R(t+1)}$ provides the solution of the dynamics. Namely,  the bulk condition cause the permutation of matrices, while applying the boundary conditions expends the support of the operator. Therefore the ansatz provides a full solution of the dynamics, provided that the initial condition is satisfied.
	}
\end{figure}

After applying the propagator, the bulk condition permutes matrices $\mA$ and $\mB$ in the bulk and changes their argument from $t$ to $t+1$, while the boundary conditions increase the support of the operator to the light-cone corresponding to the next time-step. In Figure~\ref{loc_op1} we demonstrate that if the algebraic conditions are satisfied the corresponding MPA provides the time evolution of the local operator.

\subsection{Algebraic approach for calculating physical quantities in  QHCG \label{alg_apr}}
In QHCG the algebraic relations are particularly simple and allow us to obtain explicit time independent solution for auxiliary matrices $\mA(t)=\mA$, $\mB(t)= \mB$ and boundary vectors $\bra{\mathfrak{L}(t)}=\bra{\mathfrak{L}}$ and $\ket{R(t)}=\ket{R}$, which is outlined in \ref{apA}.

In order to calculate physical quantities, we will now use an explicit form of auxiliary matrices, which tends to be quite complicated, but rather employ simple algebraic relations that these auxiliary matrices satisfy directly. Note that finding an explicit solution of the algebraic relations is nevertheless important, as it demonstrates that such solution exists and that algebraic relations can be used to calculate physical quantities. In order to demonstrate how the method works let us restrict the discussion to the observables $O=E^{zz}_1$ that are initially diagonal. As noted before the subspace of diagonal observables is preserved by the time evolution, which means that we can set all of the matrix product ansatz elements associated with the off-diagonal elements with $s\neq s'$ to $0$: $A_{s s'}=0$, $B_{s s'}=0$, and  $\bra{L_{s s'}}=0$.
The ansatz can therefore be expressed as
\begin{equation}
	\label{diag}
	E^{zz}_{1}(t)=\sum_{\underline{s}}\bra{L_{s_{-t+1}}}A_{s_{-t+2}}B_{s_{-t+3}}\cdots B_{s_{t-1}} A_{s_t}\ket{R} E^{s_{-t+1}s_{-t+1}}\dots E^{s_t s_t},
\end{equation}
where we identified the diagonal matrix product ansatz elements $ A_s\equiv A_{ss}$, $ B_s\equiv B_{ss}$, and $\bra{L_s}\equiv\bra{L_{ss}}$.
A set of bulk algebraic relations associated with diagonal elements reads
\begin{align}
	\label{alg_MPA_cl}
	A_{s} B_{s'}&=B_{s}A_{s'};\quad s,s'\neq0,\nonumber\\
	A_{s} B_{0}&=B_{0}A_{s},\\
	A_{0} B_{s}&=B_{s}A_{0},\nonumber
\end{align}
with the following boundary
\begin{align}
	\label{in_bc_cl}\nonumber
	\bra{L_s} A_{s'}&=\bra{L_{s'}}; \quad s\neq0,\nonumber\\
	\bra{L_{0}} A_{s'}&=\bra{L_{0}},\nonumber\\
	\bra{L_{s}} A_{0}&=\bra{L_{s}},\\
	B_{s} \ket{R}&=\ket{R},\nonumber
\end{align}
and initial conditions
\begin{equation}
	\label{init}
	\bra{L_{s}}R\rangle=\delta_{s,z}.
\end{equation}
We will show how to use the above algebraic relations in order to establish the weight
\begin{equation}
	\label{coef}
	\bra{L_{s_{-t+1}}}  A_{s_{-t+2}}\cdots B_{s_{t-1}} A_{s_{t}} \ket{R},
\end{equation}
of an arbitrary operator
\begin{equation}
	E^{s_{-t+1}s_{-t+1}}_{-t+1}\dots E^{s_t s_t}_{t},
\end{equation}
at time $t$ in the time evolution of observable $E^{zz}_1$.
Firstly, the bulk conditions \eqref{in_bc_cl} can be used to bring all of the auxiliary matrices $B_{s}$ in the expression \eqref{coef} to the right  of the auxiliary matrices $A_{s'}$. Matrices $B_{s'}$ can then be evaluated on the right vector $\ket{R}$ and matrices $A_{s}$ on the left vector $\bra{L_{s}}$ using the boundary conditions \eqref{in_bc_cl}. What we are finally left with is an inner product between the left and the right boundary vectors, which can be evaluated using the initial condition \eqref{init}. Calculations of more complicated objects, which we present below, rely on the same idea of commuting matrices $A$ to the left and matrices $B$ to the right using the bulk conditions, applying them to the boundary vectors, and evaluate the inner product of the boundary vectors using the initial conditions.

\section{Out-of-time ordered correlation function \label{OTOC}}
{The first quantities that we focus on are OTOCs \cite{larkin1969quasiclassical,khemani2018velocity,gopalakrishnan2018hydrodynamics,nahum2018operator,von2018operator,Chan2018Solution,gopalakrishnan2018operator,pappalardi2018scrambling,foini2019eigenstate,bernard2021dynamics,bensa2021two,lerose2020bridging}, which are in the semi-classical limit related to Lyapunov exponents \cite{maldacena2016bound}.
In the first part we will consider OTOCs between diagonal and off-diagonal observables for a generic reversible quantum cellular automaton evaluated at the infinite temperature
\begin{equation}
		C(i,t)=\frac{\text{tr}([E^{xy}_i(0),E^{zz}_1(t)][E^{xy}_i(0),E^{zz}_1(t)]^\dagger)}{\text{tr}(\one)}.
\end{equation}
In general it comprises two parts that are obtained by expanding the two commutators and taking into account that $E^{xy}E^{uv}=\delta_{yu} E^{xv}$. The first part corresponds to the time ordered two-point function
\begin{equation}
	\label{time_ordered}
	C_2(i,t)=\frac{\text{tr}(E^{yy}_i(0)E^{zz}_1(t))}{\text{tr}(\one)}+\frac{\text{tr}(E^{xx}_i(0) E^{zz}_1(t))}{\text{tr}(\one)},
\end{equation}
and the second one to the out-of-time ordered correlation function
\begin{equation}
	\label{4ptf}
	C_4(i,t)=-2\frac{\text{tr}(E^{xy}_i(0)E^{zz}_1(t)E^{yx}_i(0)E^{zz}_1(t))}{\text{tr}(\one)}.
\end{equation}
Note that the dynamics of the diagonal two point functions is governed by the classical cellular automaton time evolution that can be obtained by mapping the  diagonal observables to states $\ket{\underline{s}}\bra{\underline{s}}\to \ket{\underline{s}}$.

Due to the determinstic nature of the time evolution the four point function simplifies significantly and can be expressed in terms of pointer states as 
\begin{align}
\begin{split}
	\label{4ptf22}
	C_4(i,t)=&-\frac{2}{q^{L}}\sum_{\underline{s}^{(1)},\underline{s}^{(2)},\underline{s}^{(3)}}\bra{\cdots s^{(2)}_{i-1} x s^{(2)}_{i+1}\cdots} U(t)\ket{\cdots s^{(1)}_{0} z s^{(1)}_{2}\cdots}\times
	\\
	&\times\bra{\cdots s^{(2)}_{i-1} y s^{(2)}_{i+1}\cdots} U(t)\ket{\cdots s^{(3)}_{0} z s^{(3)}_{2}\cdots}.
\end{split}
\end{align}
Here we used $L$ for the size of the system, which should be chosen larger then the size of the light-cone at time $t$. For details see the \ref{apB}. In order to derive this expression one has to take into account that a given configuration maps  exactly into a single other configuration under the time evolution.}

If we consider a particular case of the hardcore interacting gas with $q-1$ particle species the propagator corresponds to
\begin{equation}
U(t)=
\begin{cases}
	\left(\prod_{i\in 2\ZZ} U_{ii+1}\prod_{i\in 2\ZZ+1} U_{ii+1}\right)^{t/2},\quad t\in 2\NN \\
	\prod_{i\in 2\ZZ+1} U_{ii+1}\left(\prod_{i\in 2\ZZ} U_{ii+1}\prod_{i\in 2\ZZ+1} U_{ii+1}\right)^{(t-1)/2},\quad t\in 2\NN-1
\end{cases}
\end{equation}
in terms of the local gates \eqref{lprop}.
In the following two subsections we present the matrix product ansatz based derivation of OTOCs.
\subsection{Two point correlation functions in QHCG}
In order to evaluate the two-point correlation function of diagonal operators \eqref{time_ordered}, we only need to consider the invariant subspace of diagonal operators. All of the off-diagonal entries in the ansatz from equation \eqref{ansatz} can be set to $0$, since the cellular automaton evolution preserves it, as we argued in section~\ref{alg_apr}. The problem is therefore equivalent to the calculation of the dynamical structure factors for the classical case \cite{10.21468/SciPostPhys.6.6.074}. We will evaluate here the correlation function using the algebraic approach outlined in \ref{alg_apr}. 
For simplicity we will evaluate only OTOCs for space-time coordinates that satisfy the restriction $i+t\in 2\ZZ$.
Evaluating the two point functions \eqref{time_ordered} using the matrix product ansatz \eqref{diag} for the time evolution of $E_1^{zz}(t)$ yields
\begin{equation}
	C_2(2k-t,t)=\frac{1}{\mathcal{N}}\bra{L}(T_A T_B)^{k-1}A_x(T_B T_A)^{t-k}\ket{R}+\frac{1}{\mathcal{N}}\bra{L}(T_A T_B)^{k-1}A_y(T_B T_A)^{t-k}\ket{R},
\end{equation}
in terms of transfer matrices
\begin{equation}
	T_A=\sum_{s=0}^{q-1}A_{s},\quad T_B=\sum_{s=0}^{q-1}B_{s},
\end{equation}
and boundary vectors $\ket{R}$ and
$
	\bra{L}=\sum_s\bra{L_{s}}.
$
{ $2k$ corresponds to the distance of the point in the two-point correlation function from the left boundary.
The normalization factor $\mathcal{N}$ can be determined by considering the trace of the time evolved operator and reads
\begin{equation}
\mathcal{N}=
\bra{L}(T_A T_B)^{t-1}T_A\ket{R}.
\end{equation}
Using the bulk algebraic relations \eqref{alg_MPA_cl} one can show that transfer matrices $T_A$ and $T_B$ commute, while the boundary conditions imply that $\bra{L}$ is the left eigenvector of $T_A$ and $\ket{R}$ the right eigenvector of $T_B$
\begin{equation}
	\label{relats}
	[T_A,T_B]=0,\quad\bra{L}T_A=q\bra{L},\quad T_B \ket{R}=q\ket{R}.
\end{equation} 
These relations can be used to establish the value of normalization constant  $\mathcal{N}=q^{2t-1}$, and reduce the two point correlation functions to
\begin{equation}
	\label{c222}
	C_2(2k-t,t)=q^{-t}(\bra{L}T_B^{k-1}A_x T_A^{t-k}\ket{R}+\bra{L}T_B^{k-1}A_y T_A^{t-k}\ket{R}).
\end{equation}
In order to obtain the expression above we commuted the matrices $T_A$ on the left of the operators $A_x$ and $A_y$ to the left boundary, and matrices $T_B$ to the right boundary, and subsequently used the appropriate boundary relations.
In order to evaluate this expression it proves useful to decompose transfer matrices into the part associated with vacancies and the part  $\tilde{T}_X$  corresponding to the sum over the particle contributions
\begin{equation}
 T_A=\tilde{T}_A+A_{0},\quad  T_B=\tilde{T}_B+B_{0}
\end{equation}}
 The first thing to notice is that $A_{0}$ and $B_{0}$ commute with all other auxiliary matrices, which allows us to evaluate their contribution explicitly
\begin{equation}
	C_2(2k-t,t)=q^{-t}\sum_{l_1,l_2}\binom{k-1}{l_1}\binom{t-k}{l_2}\left(\bra{L} \tilde{T}_B^{l_1}A_x\tilde{T}_A^{l_2}\ket{R}+\bra{L} \tilde{T}_B^{l_1}A_y\tilde{T}_A^{l_2}\ket{R}\right).
\end{equation}
Here $l_1$ and $l_2$ are the numbers of components $B_{s}$ and $A_{s}$, where $s\neq0$.
Accordingly to \eqref{alg_MPA_cl} the matrices associated with particles $A_{s}$ and $B_{s'}$ can be exchanged while keeping the ordering of indexes $s$ intact. This allows us to bring matrices $A_{s}$ to the left and matrices $B_{s}$ to the right and apply them to the boundary vectors accordingly to the boundary conditions \eqref{in_bc_cl}
\begin{equation}
\bra{L}\tilde{T}_A=(q-1)\bra{L},\quad \tilde{T}_B \ket{R}=(q-1)\ket{R}.
\end{equation}
Finally, after using the initial condition \eqref{init}, we obtain the expression for the two point correlation function
\begin{equation}
	C_2(2k-t,t)=\sum_{l_1,l_2}\binom{k-1}{l_1}\binom{t-k}{l_2}(q-1)^{l_1+l_2-t}(1-\delta_{l_1,l_2})+(q-1)^{l_1+l_2-t+1}\delta_{l_1,l_2}(\delta_{z,x}+\delta_{z,y}),
\end{equation}
which can be simplified to
\begin{equation}
	\label{two_point}
	C_2(2k-t,t)=2 q^{-1}+\frac{(q-1)(\delta_{z,x}+\delta_{z,y})+2}{q}\sum_l \binom{k-1}{l}\binom{t-k}{l}(1-1/q)^{2l}q^{-(t-2l-1)},
\end{equation}
by employing the identity
\begin{equation}
	\sum_{l_1,l_2}\binom{k-1}{l_1}\binom{t-k}{l_2}(q-1)^{(l_1+l_2)}=q^{(t-1)}.
\end{equation}

\subsection{Four point correlation functions in  QHCG}
In order to evaluate the four point correlation function \eqref{4ptf22} we can again employ the ansatz \eqref{diag}. The central observation is that the time evolution of $\sum_{\underline{s}'}\ket{\cdots s'_{0} z s'_{2}\cdots}$ is encoded by the ansatz for diagonal operators \eqref{diag}
\begin{equation}
	U(t)\left(\sum_{\underline{s}'}\ket{\cdots s'_{0} z s'_{2}\cdots}\right)=\sum_{\underline{s}'}\bra{L_{s_{-t+1}'}}A_{s_{-t+2}'}B_{s_{-t+3}'}\cdots B_{s_{t-1}'} A_{s_t'}\ket{R} \ket{\cdots s'_{0} s'_{1}. s'_{2}\cdots}.
\end{equation}
This correspondence is established by identifying the diagonal operator $E^{ss}$ with the state $\ket{s}$, and allows us to express the correlation function \eqref{4ptf22} in the matrix product form
\begin{align}
	\begin{split}
	C_4(2k-t,t)&=\sum_{\ul{s}}\bra{L_{s_{-t+1}}}A_{s_{-t+2}}B_{s_{-t+3}}\cdots \underbrace{A_{x}}_{\text{site}\ i}\cdots B_{s_{t-1}}A_{s_{t}}\ket{R}\times\\
	&\times
	\bra{L_{s_{-t+1}}}A_{s_{-t+2}}B_{s_{-t+3}}\cdots \underbrace{A_{y}}_{\text{site}\ i}\cdots B_{s_{t-1}} A_{s_{t}}\ket{R}.
	\end{split}
\end{align}
After introducing the doubled transfer matrices and boundary vectors 
\begin{align}
	\begin{split}
	\mathbb{T}_A&=\sum_r A_r\otimes A_r,\\
	\mathbb{T}_B&=\sum_r B_r\otimes B_r,\\
	\bra{\mathbb{L}}&=\sum_r \bra{L_r}\otimes\bra{L_r},\\
	\ket{\mathbb{R}}&=\ket{R}\otimes\ket{R},
	\end{split}
\end{align}
 we get the expression for the four point function
\begin{equation}
	C_4(2k-t,t)=-\frac{2}{\mathcal{N}_2}\bra{\mathbb{L}}(\mathbb{T}_A\mathbb{T}_B)^{k-1}A_x\otimes A_y(\mathbb{T}_B\mathbb{T}_A)^{t-k}\ket{\mathbb{R}}.
\end{equation}
{Again the normalization is determined by
\begin{equation}
	\mathcal{N}_2=
	\bra{\mathbb{L}}(\mathbb{T}_A \mathbb{T}_B)^{t-1}\mathbb{T}_A\ket{\mathbb{R}}.
\end{equation}
We evaluate this expression using the algebraic relations
\begin{equation}
	[\mathbb{T}_A,\mathbb{T}_B]=0,\quad\bra{\mathbb{L}}\mathbb{T}_A=q\bra{\mathbb{L}},\quad \mathbb{T}_B \ket{\mathbb{R}}=q\ket{\mathbb{R}},
\end{equation} 
which follow from the bulk and boundary conditions \eqref{alg_MPA_cl}. Commutativity of doubled transfer matrices yields
\begin{equation}
		C_4(2k-t,t)=-2 q^{-t}\bra{\mathbb{L}}\mathbb{T}_B^{k-1}A_x\otimes A_y \mathbb{T}_A^{t-k}\ket{\mathbb{R}}.
\end{equation}
The vacancy component associated with the doubled transfer matrices is $A_0\otimes A_0$ and therefore again commute with all MPA entries, while the order of doubled auxiliary matrices associated with particles can be exchanged and applied to the boundary vectors.} The four point function therefore reads
\begin{equation}
	\label{four_point}
	C_4(2k-t,t)=-2q^{-1}-2\frac{(q-1)\delta_{z,x}\delta_{z,y}-1}{q}\sum_l \binom{k-1}{l}\binom{t-k}{l}(1-1/q)^{2l}q^{-(t-2l-1)}.
\end{equation}

Putting together the expressions for two point \eqref{two_point} and four point \eqref{four_point} functions we get the expression for the out-of-time ordered correlation function
\begin{equation}
	\label{OTOC_f}
	C(i,t)=\frac{(q-1)(\delta_{z,x}+\delta_{z,y}-2\delta_{z,x}\delta_{z,y})}{q}\sum_l^{\text{min}(k-1,t-k)} \binom{k-1}{l}\binom{t-k}{l}(1-1/q)^{2l}q^{-(t-2l-1)}.
\end{equation}
Note that OTOCs vanish except if either $x=z$ or $y=z$.
In this case, if we take the limit of infinite number of particle species $q\to\infty$, we see that OTOCs increase linearly with time on site $0$ and site $2$
\begin{equation}
	\label{asymp}
	\lim_{q\to\infty} q \times C(i,t)= \begin{cases}
		\frac{t}{2}; & i=\{0,2\}, \\
		0; & \text{Otherwise}.
	\end{cases}
\end{equation}
We have therefore shown that in the limit of the infinite local Hilbert space dimension OTOCs can exhibit unbounded increase even in integrable systems.
Similar behavior was observed in chaotic quantum systems with a finite local Hilbert space dimension, for extensive operators, and was dubbed weak quantum chaos \cite{kukuljan2017weak}.

\begin{figure}
	\centering
	\includegraphics[width=0.5\textwidth]{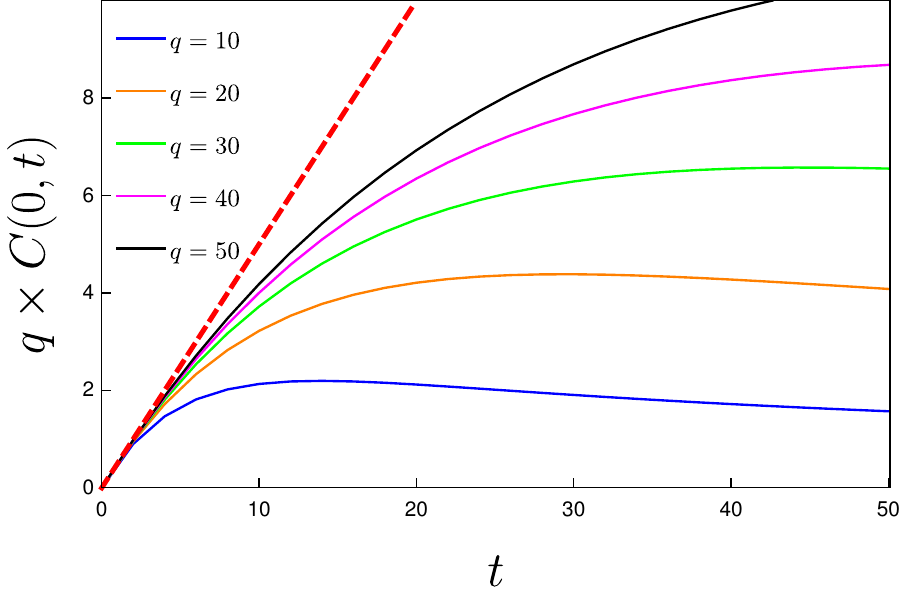}\includegraphics[width=0.5\textwidth]{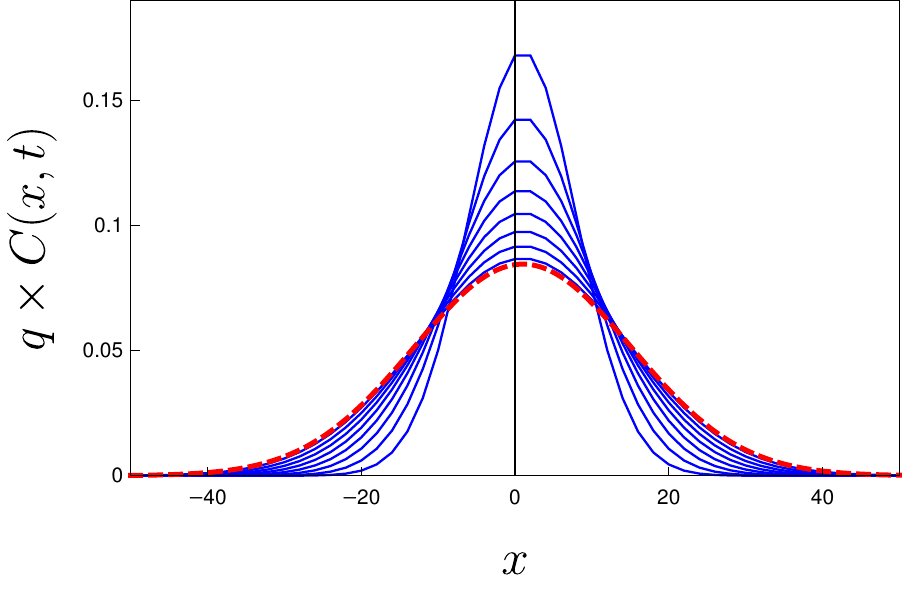}
	\caption{\label{fig:OTOC} The dynamics of the OTOCs for different local Hilbert space dimension $q$  on the left. The red dashed line denotes the asymptotic, $q\to\infty$, behavior \eqref{asymp}. On the right we plot the profile of the OTOC between $t=100$ and $t=400$ for $q=3$. The red dashed line corresponds to the continuum approximation \eqref{cont_OTOC} at $t=400$.
	}
\end{figure}

Finally, considering the long time limit of expression \eqref{OTOC_f}, we can establish that it converges to the Gaussian \cite{10.21468/SciPostPhys.6.6.074}
\begin{equation}
	\label{cont_OTOC}
	\lim_{t\to\infty}C(x\sqrt{t},t)=\frac{1-1/q}{\sqrt{2\pi t/q(1-1/q)}} \exp\left(-\frac{(q-1)x^2}{2}\right).
\end{equation}
In Figure~\ref{fig:OTOC} we plot $C$ for different values of $q$ and compare it to the asymptotic behavior \eqref{asymp}.
\section{Operator  weight spreading \label{OW}}
The spreading of the operator weight quantifies how the size of operators changes with time. { In \cite{lopez2021operator} it was conjectured that the operator front can be used to distinguish between the integrable and non-integrable dynamics. Furthermore in \cite{von2018operator} it was shown that in random unitary circuits operator weight spreading relates to OTOCs.}
When considering the operator weight spreading it is more convenient to deal with orthogonal local operator basis $\{\kappa_i\}$, such that
\begin{equation}
	\kappa_0=\one, \quad \text{tr}(\kappa_i\kappa_j)=q\,\delta_{i,j},\quad \kappa_i^\dagger=\kappa_i,
\end{equation}
instead of the operators $E^{kl}$. Let us remind the reader that $q$ is the local Hilbert space dimension.
The right operator weight $\rho_R(i,t)$ of the operator
\begin{equation}
	\label{op}
	O(t)=\sum_{\ul{s}}a_{\cdots s_{-1}s_0 s_1\cdots}(t) \cdots\otimes\kappa_{s_{-1}}\otimes \kappa_{s_{0}}\otimes \kappa_{s_{1}}\otimes \cdots
\end{equation}
 corresponds to the weight of all operators in the expansion \eqref{op}, for which the support on the right corresponds to the coordinate $i$\footnote{This means that after the point $i$ the operators act like an identity.}, 
 {
\begin{equation}
	\rho_R(i,t)=\sum_{i=\text{max} (x) \text{; such that} \ s_x\neq 0} |a_{\cdots s_{-1}s_0 s_1\cdots}(t)|^2.	
\end{equation}
$\rho_R(i,t)$ can be understood as the density, since the summation over all end points of the operators $i$ in the expansion \eqref{op} corresponds to the square of the Hilbert-Schmidt norm of the time evolved operator $O(t)$, which is preserved in time
\begin{equation}
	\sum_i \rho_R(i,t)=\text{tr}(O(t)O(t)^\dagger)=\text{tr}(OO^\dagger).
\end{equation}
In order to establish the form of the front we will again use the MPA representation in terms of matrices $C$ in the local operator basis $\kappa_s$}. In general, if the operator is supported on the finite sub-lattice $[-t+1,t]$
\begin{equation}
	O=\sum_{\ul{s}}\bra{L_{s_{-t+1}}}C_{s_{-t+2}}C_{s_{-t+3}}\cdots C_{s_{t-1}} C_{s_t}\ket{R} \cdots\kappa_0 \otimes\kappa_{s_{-t+1}}\otimes\cdots\otimes \kappa_{s_{t}}\otimes\kappa_0 \cdots,
\end{equation}
the right density {in the MPA language} reads
\begin{equation}
	\label{right_density}
	\rho_R(i,t)=\frac{1}{q^{2t}}\bra{  \mathbb{L}}\mathbb{T}^{t-2+i}\mathbb{X}\mathbb{T}_\one^{t-i}\ket{  \mathbb{R}},
\end{equation}
for $i\in [-t+2,t]$, where we introduced the doubled auxiliary matrices
\begin{equation}
	\bra{  \mathbb{L}}= \sum_s\bra{ L_s}\otimes\overline{\bra{ L_s}}, \quad \ket{  \mathbb{R}}= \ket{ R}\otimes \overline{\ket{ R}},
\end{equation}
\begin{equation}
	\mathbb{X}=\sum_{s\neq 0} C_s\otimes \overline{C}_s,\quad \mathbb{T}=\sum_{s} C_s\otimes \overline{C}_s,\quad \mathbb{T}_{\one}=C_0\otimes \overline{C}_0.
\end{equation}
Note that the auxiliary matrix $C_0$ in this basis is associated with the identity and should not be confused with auxiliary matrices $A_0$ or $B_0$, which correspond to the projectors onto the vacancy state.
Since $\mathbb{T}=\mathbb{T}_{\one}+\mathbb{X}$, we can alternatively represent \eqref{right_density} in terms of two operators $V_R(i,t)$ \cite{lopez2021operator}
\begin{equation}
	\rho_R(i,t)=\frac{1}{q^{2t}}\bra{  \mathbb{L}}\mathbb{T}^{t-1+i}\mathbb{T}_\one^{t-i}\ket{  \mathbb{R}}-\frac{1}{q^{2t}}\bra{  \mathbb{L}}\mathbb{T}^{t-2+i}\mathbb{T}_\one^{t-i+1}\ket{  \mathbb{R}}=V_R(i-1,t)-V_R(i-2,t).
\end{equation}

\subsection{Operator weight spreading in  QHCG}
Since we will study the operator spreading in QHCG for both the diagonal and the off-diagonal operators we have to consider the ansatz, {which includes the off-diagonal terms}. We will focus on the time evolution of the operator front of real operators, which can be evaluated using the matrices
\begin{align}
	\begin{split}
	&\mathbb{T}_A=\frac{1}{q}\sum_{s_1,s_2}A_{s_1 s_2}\otimes A_{s_1s_2}, \quad \mathbb{T}_B=\frac{1}{q}\sum_{s_1,s_2}B_{s_1 s_2}\otimes B_{s_1s_2}, \quad
	\\
	&
	A_{\one}=\frac{1}{q}\sum_s A_{ss},\quad B_{\one}=\frac{1}{q}\sum_s B_{ss},\\
	&\bra{  \mathbb{L}}=\frac{1}{q}\sum_{s_1,s_2}\bra{L_{s_1 s_2}}\otimes \bra{L_{s_1s_2}},\quad \bra{L_{\one}}=\frac{1}{q}\sum_s \bra{L}_{ss}.
	\end{split}
\end{align}
For simplicity we will consider the right operator front density averaged over two consecutive sites
\begin{align}
	\begin{split}
	\overline{\rho}_R(2k-t,t)=&\frac{1}{2}\left(\rho(2k-t+1,t)+\rho(2k-t,t)\right)=\\&=\frac{1}{2\mathcal{N}_3}\bra{\mathbb{L}}(\mathbb{T}_A\mathbb{T}_B)^{k} (A_\one\otimes A_\one B_\one\otimes B_\one)^{t-k}\ket{\mathbb{R}}-\\&-\frac{1}{2\mathcal{N}_3}\bra{\mathbb{L}}(\mathbb{T}_A\mathbb{T}_B)^{k-1} (A_\one\otimes A_\one B_\one\otimes B_\one)^{t-k+1}\ket{\mathbb{R}}.
	\end{split}
\end{align}
Note that in the above expression we added a set of operators $B_\one\otimes B_\one$ to the right, since they act trivially on the right vector $\ket{\mathbb{R}}$.
In case of  QHCG  the operators $V$ correspond to
\begin{equation}
	\label{ver}
	V_R(2k-t+1,t)=\frac{1}{\mathcal{N}_3}\bra{\mathbb{L}}(\mathbb{T}_A\mathbb{T}_B)^{k} (A_\one\otimes A_\one B_\one\otimes B_\one)^{t-k}\ket{\mathbb{R}}.
\end{equation} 
The algebraic relations which are central for evaluating  expression \eqref{ver} {and can again be derived from the bulk and boundary conditions} are
\begin{align}
	\begin{split}
	[\mathbb{T}_A,\mathbb{T}_B]&=0,\\
	\bra{\mathbb{L}}\mathbb{T}_A&= \bra{\mathbb{L}},\\
 B_{\one}\otimes B_{\one}\ket{\mathbb{R}}&=\ket{\mathbb{R}},\\
   	\ [A_\one,B_\one] &=0.
   	\end{split}
\end{align}
Employing the same strategy as in the preceding sections to evaluate the norm, we get $\mathcal{N}_3=\|O\|_{HS}^2=1/q$, 
\begin{equation}
	V_R(2k-t+1,t)=\bra{\mathbb{L}}\mathbb{T}_B^{k} (A_\one\otimes A_\one)^{t-k}\ket{\mathbb{R}}.
\end{equation}
After separating the vacancy components from others in the same fashion as in the case of the two and the four point functions, and accounting for its commutativity with all the auxiliary matrices, $V$ reduces to
\begin{equation}
	V_R(2k-t+1,t)=\sum_{l_1,l_2,l_3}\binom{k}{l_1}\binom{t-k}{l_2}\binom{t-k}{l_3}q^{k-2t+l_1+k_2+l_3}\bra{\mathbb{L}}\tilde{\mathbb{T}}_B^{l_1} (\tilde{A}_\one\otimes\one)^{l_2} (\one\otimes\tilde{A}_\one)^{l_3}\ket{\mathbb{R}}.
\end{equation}
For simplicity we will discard initial operators associated with vacancies and instead focus separately on diagonal and off diagonal terms involving only particles $E^{pp}_1(t)$ and $E^{pp'}_1(t)$, for $p'\neq p\neq0$. This allows us to set $\bra{L}_{0p}=\bra{L}_{p0}=\bra{L}_{00}=0$. Using the algebraic relations
\begin{align}
	\begin{split}
	\bra{\mathbb{L}}\tilde{\mathbb{T}}_A&=(1-1/q)\bra{\mathbb{L}},\\
	\tilde{\mathbb{T}}_B\ket{\mbb{R}}&=(\tilde{B}_\one\otimes\one)\ket{\mbb{R}}=(\one\otimes \tilde{B}_{\one})\ket{\mbb{R}}=(1-1/q)\ket{\mbb{R}},\\
	\tilde{A}_\one\otimes\one\ket{\mbb{R}}&=\frac{1}{q}\sum_{s_1,s_2\neq 0}A_{s_1s_2}\otimes B_{s_1s_2}\ket{\mbb{R}},\\
	\bra{\mbb{L}}\tilde{A}_\one\otimes\one&=\bra{L_\one}\otimes\bra{L_\one}\\
	\bra{L_\one}\otimes\bra{L_\one}\tilde{A}_\one\otimes\one&=\bra{L_\one}\otimes\bra{L_\one}\one\otimes \tilde{A}_\one=(1-1/q)\bra{L_\one}\otimes\bra{L_\one},
	\end{split}
\end{align}
which can be derived using the bulk and boundary algebraic conditions,
we can evaluate the matrix elements explicitly. If $l_1\geq l_2=l_3$ we have
\begin{equation}
	\bra{\mathbb{L}}\tilde{\mathbb{T}}_B^{l_1} (\tilde{A}_\one\otimes\tilde{A}_\one)^{l_2}\ket{\mathbb{R}}=(1-1/q)^{l_1+2l_2} \bra{\mbb{L}}\mbb{R}\rangle,
\end{equation}
and
\begin{equation}
	\bra{\mathbb{L}}\tilde{\mathbb{T}}_B^{l_1} (\tilde{A}_\one\otimes\one)^{l_2} (\one\otimes\tilde{A}_\one)^{l_3}\ket{\mathbb{R}}=(1-1/q)^{l_1+l_2+l_3-1} \bra{L_\one}\otimes \bra{L_\one}\mbb{R}\rangle,
\end{equation}
if this is not the case.
In case of diagonal operators the expression for $V$ evaluates to
\begin{align}
	\label{R1}
	\begin{split}
	V_R(i,t)&=\frac{1}{q^{(2t-k+1)}}\left(1+\sum_{l_1=1}^k\sum_{l_2=0}^{l_1}\binom{k}{l_1}\binom{t-k}{l_2}^2(q-1)^{2l_2+l_1}+\right.\\&+\sum_{l_2=1}^{t-k}\sum_{l_1=0}^{l_2-1} \binom{k}{l_1}\binom{t-k}{l_2}^2 (q-1)^{2 l_2+l_1-1}+\\&+\left.\sum_{l_1,l_2,l_3;l_1+l_2+l_3>0}(1-\delta_{l_1,l_2})\binom{t-k}{l_1}\binom{t-k}{l_2}\binom{k}{l_3}(q-1)^{l_1+l_2+l_3-1}\right),
	\end{split}
\end{align}
and
\begin{eqnarray}
	\label{R2}
	V_R(i,t)&=&\frac{1}{q^{(2t-k+1)}}\left(1+\sum_{l_1=1}^k\sum_{l_2=0}^{l_1}\binom{k}{l_1}\binom{t-k}{l_2}^2(q-1)^{2l_2+l_1}\right),
\end{eqnarray}
for off-diagonal operators, {where $i=2k-t$}. On the left and the right hand side of the light cone $V_R$ takes the values $\frac{1}{q^2}$ and $\frac{1}{q}$ respectively for diagonal operators and  $0$ and $\frac{1}{q}$ for off diagonal operators.

\begin{figure}
	\centering
	\includegraphics[width=0.5\textwidth]{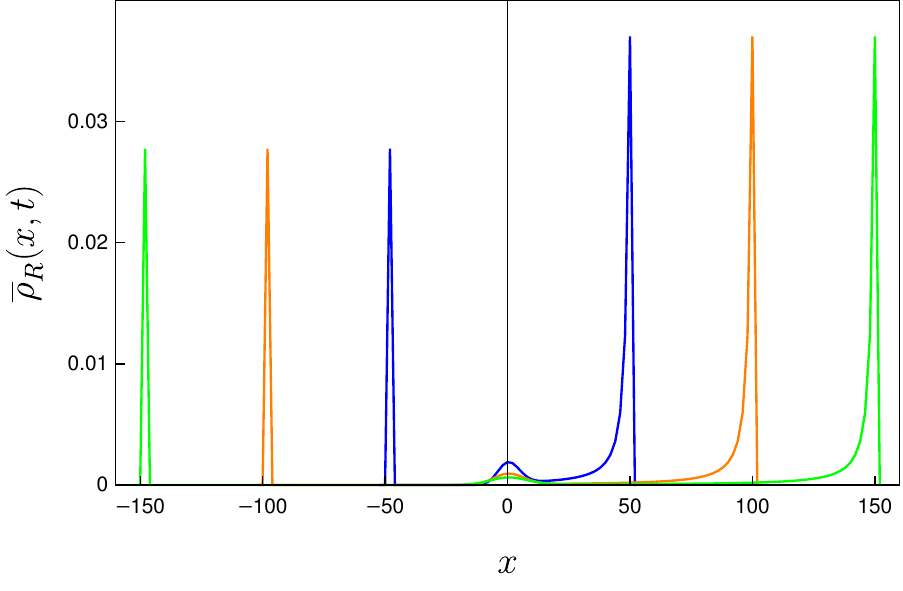}\includegraphics[width=0.5\textwidth]{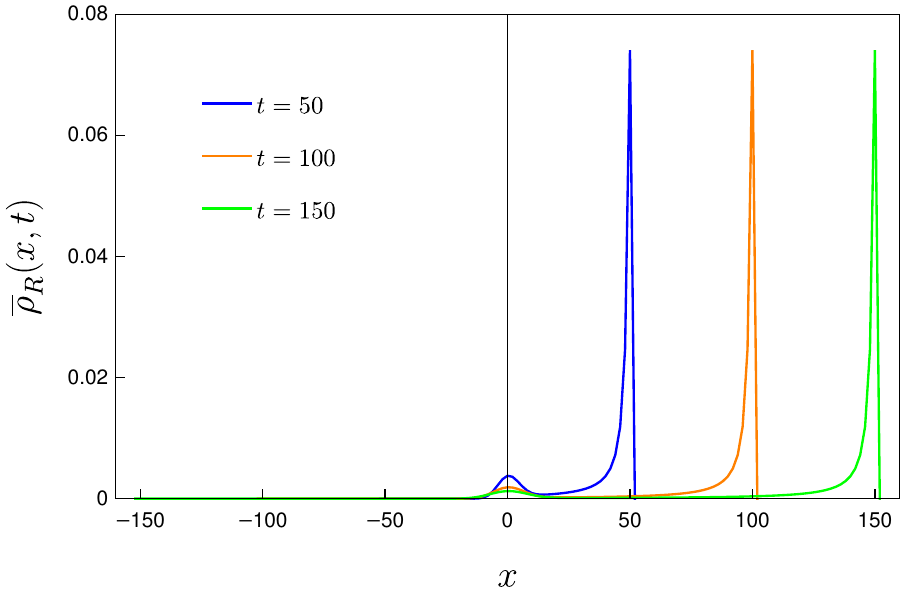}
	\caption{\label{fig:ow1} Operator weight dynamics for diagonal operators on the left and for off diagonal operators on right.
	}
\end{figure}

\begin{figure}
	\centering
	\includegraphics[width=0.5\textwidth]{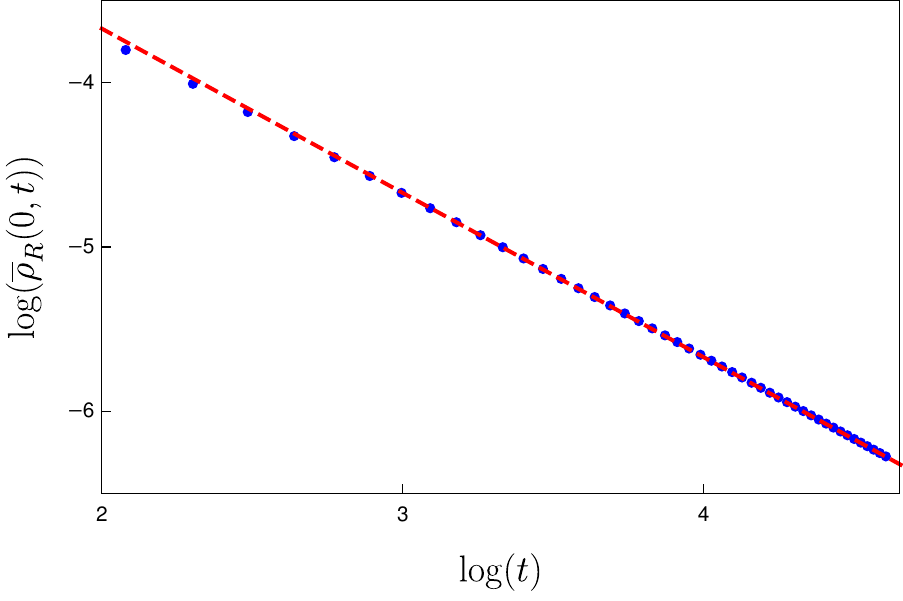}\includegraphics[width=0.5\textwidth]{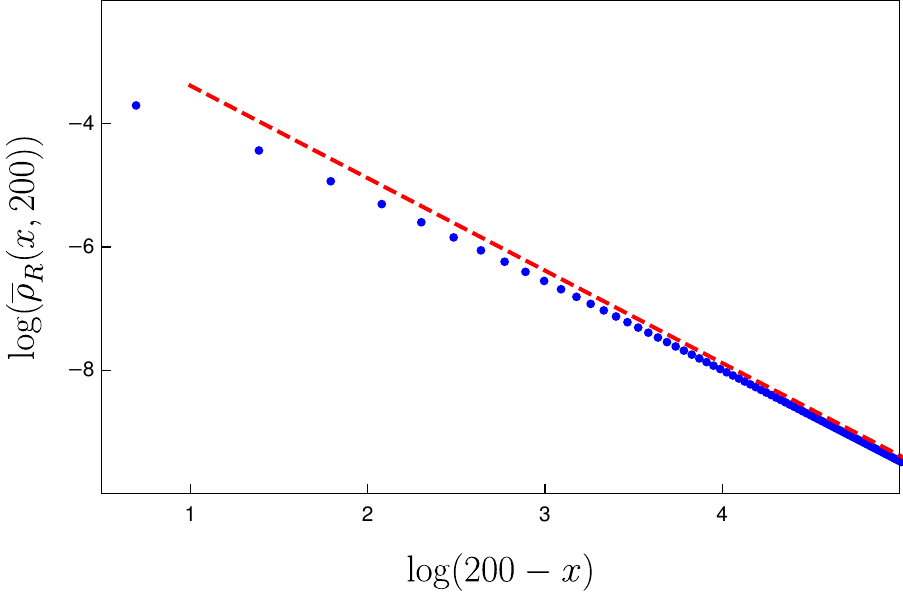}
	\caption{\label{fig:ow2} Time dependence of the central peak for off diagonal operators on the left figure. The red dashed line corresponds to $\propto1/t$ decay. Dependence of the operator front on coordinate $x$ at the right edge for off diagonal operators. The red dashed line corresponds to the $\propto 1/(t-x)^{3/2}$ scaling.
	}
\end{figure}

\begin{figure}
	\centering
	\includegraphics[width=0.5\textwidth]{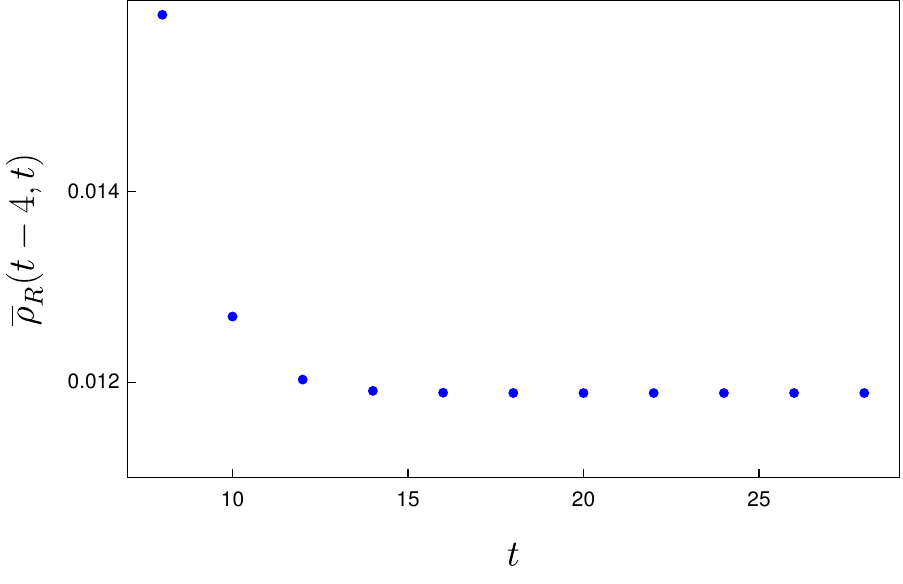}
	\caption{\label{fig:ow3} Localization of the operator front close to the right edge for off diagonal operators.
	}
\end{figure}

In Figure~\ref{fig:ow1} we plot the dynamics of the right operator weight for diagonal and off diagonal operators. We observe that the operator weight propagates ballistically with the maximal velocity $1$ and remains localized for arbitrarily long times. In  Figure~\ref{fig:ow2} on the left we show that the central peak of the operator front decays as $1/t$, and in the Figure~\ref{fig:ow2} on the right that the front scales as $1/(t-x)^{3/2}$ from the right edge which is moving ballistically with velocity 1. Finally, in Figure~\ref{fig:ow3} we show the convergence of the operator weight close to the edge of the operator front.
\section{Operator Entanglement in QHCG \label{opent}}
Finally, we will shortly discuss the time evolution of the operator space entanglement entropy in QHCG and show that its growth is  bounded by $\log(t)$, which agrees with the predictions for the behaviour of interacting integrable systems \cite{Alba2019Operator}. The upper bound is intimately connected to the fact that the computation complexity of the matrix product ansatz increases only algebraically in time. 

OSEE measures how "complicated" the decomposition of the operator into two subsystems is. Let us assume that we divide the system into two parts $A$ and $B$, and make a corresponding decomposition of the operator
\begin{equation}
	O/\sqrt{\Tr(O^\dagger O)}=\sum_{i,j} M_{i,j} O_{A,i}O_{B,j}.
\end{equation}
Operators $O_{A,i}$ and $O_{B,i}$ are supported on subsystems $A$ and $B$ and orthonormalized
\begin{equation}
\Tr_{A/B}\big(O_{A/B,i}^\dagger O_{A/B,j}\big)=\delta_{ij}\ .
\end{equation}
Finding a singular value decomposition of the matrix $[M]_{i,j}=M_{i,j}$ allows us to represent the operator $O$ as
\begin{equation}
	O/\sqrt{\Tr(O^\dagger O)}=\sum_{i}\sqrt{\lambda_i}O_{A,i} O_{B,i}.
\end{equation}
Schmidt coefficients $\lambda_i$ are normalized,  $\sum_i\lambda_i=1$, and the Von Neumann operator space entanglement entropy corresponds to
\begin{equation}
	S=-\sum_i\lambda_i\log \lambda_i.
\end{equation} 
OSEE is naturally related to the complexity of the operator decomposition, which can be demonstrated by focusing on two marginal cases. If the operator corresponds to the tensor product of two operators, one that is supported on the sub-lattice $A$ and the other one on the sub-lattice $B$, then the OSEE is $0$, since there is only a single coefficient $\lambda_i=1$. If, on the other hand, the weights of the operator are equivalently distributed between all the orthogonal operators in the smaller subsystem and corresponding orthogonal operators on the larger subsystem then OSEE is maximal and $$\lambda_i=\frac{1}{\text{Dim.}},$$
where $\text{Dim.}$ is the dimension of the operatorial subspace $A$.
In the case of the QHCG the restriction of the dynamics to the light-cone immediately implies that OSEE for the operators that are localized at time $t=0$ is upper bounded by
\begin{equation}
	S\leq 2t \log q,
\end{equation}
since $q^{2t}$ is the operator space dimension of the equivalent bipartition inside of the light-cone.
Matrix product ansatz, which is presented in \ref{apA}, provides a much tighter bound for bipartitions $A$ and $B$ that are connected in space. In general it can be shown that an upper bound on the operator entanglement in terms of the dimension Dim. of MPA exists \cite{Alba2019Operator}
\begin{equation}
	S\leq \log(\text{Dim.}).
\end{equation}
The dimension of the MPA \eqref{sol} is in general infinite. However, the infinite dimensional operators in \eqref{auxop} can be truncated to the dimension which increases linearly with time due to the form of the boundary operators \eqref{bv}. This implies that the computational cost for representing an operator with the MPA also increases only linearly with time, providing a logarithmic bound on OSEE
\begin{equation}
		S\leq \text{Const.}\times \log(t).
\end{equation}

\section{Conclusion}
In this article we calculated three different quantities that are associated with operator spreading in the set of integrable quantum cellular automata, which are expected to distinguish between integrable and chaotic dynamics. Firstly, we have shown that OTOCs exhibit linear increase with time in the limit of the infinite local Hilbert space dimension, despite the integrability of the model. Such behaviour was dubbed weak quantum chaos in \cite{kukuljan2017weak}. Operator front, on the other hand, travels ballistically with the maximal velocity, while the shape of the profile freezes in the long time limit as shown in the Figures~\ref{fig:ow1} \ref{fig:ow3}, which goes against what is expected in generic integrable systems \cite{lopez2021operator}. The final quantity that we discuss is the operator space entanglement entropy. We show that in the case of QHCG the increase of OSEE is upper bounded by the logarithmic increase in time, in accordance with predictions in \cite{Alba2019Operator}.

The question remains whether more complicated integrable models with infinite number of local degrees of freedom can exhibit exponential increase of OTOCs, and what is the nature of the operator weight spreading in generic integrable systems.

In order to evaluate the physical quantities, such as OTOCs and the operator front spreading we used the algebraic approach, which does not rely on the explicit matrix product ansatz representation of the dynamics, contrary to previous works, but rather employs algebraic conditions directly. This makes calculations of physical quantities much simpler. 

A future perspective is to understand for what kind of local gates $U$ the algebraic conditions \ref{cond} can be solved. In principle they could provide an algebraic way to solve the dynamics of a vast range of quantum systems exactly, or simulate their late time dynamics efficiently by finding the approximate solutions.

\section*{Acknowledgements}
This work was supported by the European Research Council (ERC) under the European Union's Horizon 2020 research and innovation programme (grant agreement No. 864597) and the FNS/SNF Ambizione Grant PZ00P2$\_$202106.
I would like to thank V. Alba, J. Dubail, K. Klobas, T. Prosen and M. Vanicat for collaborations on related topics.

\appendix
\section{Four point function for deterministic systems \label{apB}}
In order to obtain the expression \eqref{4ptf22} in the main text we insert a resolution of operators $E^{xz}$ and $E^{zz}$ in \eqref{4ptf22} in terms of the pointer states
\begin{align}
	\begin{split}
		\label{4pt}
		C_4(i,t)=-\frac{2}{q^{L}}\sum_{\underline{s}^{(1)},\underline{s}^{(2)},\underline{s}^{(3)},\underline{s}^{(4)}}&\bra{\cdots s^{(4)}_{i-1} x s^{(4)}_{i+1}\cdots} U(t)\ket{\cdots s^{(1)}_{0} z s^{(1)}_{2}\cdots}\times\\&\times\bra{\cdots s^{(1)}_{0} z s^{(1)}_{2}\cdots}U(t)^\dagger\ket{\cdots s^{(2)}_{i-1} x s^{(2)}_{i+1}\cdots}\times\\
		&\times\bra{\cdots s^{(2)}_{i-1} y s^{(2)}_{i+1}\cdots} U(t)\ket{\cdots s^{(3)}_{0} z s^{(3)}_{2}\cdots}\times\\&\times\bra{\cdots s^{(3)}_{0} z s^{(3)}_{2}\cdots}U(t)^\dagger\ket{\cdots s^{(4)}_{i-1} y s^{(4)}_{i+1}\cdots},
	\end{split}
\end{align}
where $L$ is the size of the system, which is larger then the light-cone at time $t$. Due to the bijective nature of the determinstic evolution we can have nonzero contribution to the four point function \eqref{4pt} only if $s^{(4)}_k=s^{(2)}_k$, since the state $\bra{\cdots s^{(1)}_{0} z s^{(1)}_{2}\cdots}$ cannot be mapped to two distinct states. This means that the four point function simplifies to 
\begin{eqnarray*}
	C_4(i,t)=-\frac{2}{q^{L}}\sum_{\underline{s}^{(1)},\underline{s}^{(2)},\underline{s}^{(3)}}\bra{\cdots s^{(2)}_{i-1} x s^{(2)}_{i+1}\cdots} U(t)\ket{\cdots s^{(1)}_{0} z s^{(1)}_{2}\cdots}^2\times
	\\
	\times\bra{\cdots s^{(2)}_{i-1} y s^{(2)}_{i+1}\cdots} U(t)\ket{\cdots s^{(3)}_{0} z s^{(3)}_{2}\cdots}^2.
\end{eqnarray*}
Again, due to the deterministic evolution the matrix elements 
\begin{equation}
\bra{\cdots s^{(2)}_{i-1} x s^{(2)}_{i+1}\cdots} U(t)\ket{\cdots s^{(1)}_{0} z s^{(1)}_{2}\cdots}
\end{equation}
can only take the values $0$ or $1$, which means that we can drop the squares in the above expression obtaining the result
\begin{eqnarray*}
	C_4(i,t)=-\frac{2}{q^{L}}\sum_{\underline{s}^{(1)},\underline{s}^{(2)},\underline{s}^{(3)}}\bra{\cdots s^{(2)}_{i-1} x s^{(2)}_{i+1}\cdots} U(t)\ket{\cdots s^{(1)}_{0} z s^{(1)}_{2}\cdots}\times
	\\
	\times\bra{\cdots s^{(2)}_{i-1} y s^{(2)}_{i+1}\cdots} U(t)\ket{\cdots s^{(3)}_{0} z s^{(3)}_{2}\cdots}.
\end{eqnarray*}
\section{tMPA for QHCG \label{apA}}
In this appendix we will outline the explicit tMPA solution of QHCG. The auxiliary matrices $A_{p_1 p_2}$ and $B_{p_1' p_2'}$ associated with particles $p_{i}\neq 0$, $p_{i}'\neq 0$ should satisfy the relations 
\begin{eqnarray}
	{ {A}}_{p_1p_2}{ {B}}_{p_1'p_2'}={ {B}}_{p_1p_2}{ {A}}_{p_1'p_2'}.
\end{eqnarray}
Relations for other components read
\begin{align}
	\begin{split}
	{ {A}}_{0 0}{ {B}}_{p_1'p_2'}&={ {B}}_{p_1'p_2'}{ {A}}_{00},\\
	{ {A}}_{p_1 p_2}{ {B}}_{00}&={ {B}}_{00}{ {A}}_{p_1p_2},\\
	{ {A}}_{00}{ {B}}_{0 0}&={ {B}}_{0 0}{ {A}}_{00}.
	\end{split}
\end{align}
\begin{align}
	\label{rel_odd}
	\begin{split}
		{ {A}}_{0 p_2}{ {B}}_{p_1'p_2'}&={ {B}}_{p_1'p_2}{ {A}}_{0p_2'},\\
		{ {A}}_{p_1 0}{ {B}}_{p_1'p_2'}&={ {B}}_{p_1p_2'}{ {A}}_{p_1'0},\\
		{ {A}}_{p_1 p_2}{ {B}}_{0p_2'}&={ {B}}_{0p_2'}{ {A}}_{p_1p_2'},\\
		{ {A}}_{p_1 p_2}{ {B}}_{p_1'0}&={ {B}}_{p_10}{ {A}}_{p_1'p_2},\\
		{ {A}}_{p_1 0}{ {B}}_{0p_2'}&={ {B}}_{0p_2'}{ {A}}_{p_10},\\
		{ {A}}_{p_1 p_2}{ {B}}_{00}&={ {B}}_{00}{ {A}}_{p_1p_2},\\
		{ {A}}_{0 p_2}{ {B}}_{p_1'0}&={ {B}}_{p_1'0}{ {A}}_{0p_2},\\
		{ {A}}_{p_1 0}{ {B}}_{p_1'0}&={ {B}}_{p_1 0}{ {A}}_{p_1'0},\\
		{ {A}}_{0 p_2}{ {B}}_{0 p_2'}&={ {B}}_{0 p_2}{ {A}}_{0 p_2'},\\
		{ {A}}_{0 0}{ {B}}_{0 p_2'}&={ {B}}_{0 p_2'}{ {A}}_{00},\\
		{ {A}}_{0 0}{ {B}}_{p_1' 0}&={ {B}}_{p_1' 0}{ {A}}_{00},\\
		{ {A}}_{0 p_2}{ {B}}_{0 0}&={ {B}}_{0 0}{ {A}}_{0p_2},\\
		{ {A}}_{p_1 0}{ {B}}_{0 0}&={ {B}}_{0 0}{ {A}}_{p_10}.
	\end{split}
\end{align}
We can satisfy relations \eqref{rel_odd} by choosing
\begin{equation}
	{  A }_{0,p}={  A }_{p,0}={  B }_{0,p}={  B }_{p,0}=0, \quad p\neq 0.
\end{equation} 
This is a consequence of ballistic propagation of vacancies $0$, which means that combinations $\ket{0}\bra{p}$ and $\ket{p}\bra{0}$ can only appear at the edge of the light-cone. Such operators are therefore fully captured by specifying appropriate components of boundary vectors $\bra{  L}_{0p}$ and $\bra{  L}_{p0}$. Boundary conditions which the ansatz should satisfy are
\begin{equation}
	{  B}_{p_1 p_2}\ket{ {R}}=\delta_{p_1,p_2} \ket{ {R}},
\end{equation}
\begin{align}
\begin{split}
		\bra{  L_{p_1 p_2}}{  A}_{p_3 p_4}&=\delta_{p_1 p_2}\bra{  L_{p_3 p_4}},\\
		\bra{  L_{0 0}}{  A}_{p_3 p_4}&=\delta_{p_3 p_4}\bra{  L_{00}},\\
		\bra{  L_{p_1 p_2}}{  A}_{0 0}&=\bra{  L_{p_1 p_2}},\\
		\bra{  L_{p_1 0}}{  A}_{p_3 p_4}&=\delta_{p_1 p_4}\bra{  L_{p_3 0}},\\
		\bra{  L_{0 p_2}}{  A}_{p_3 p_4}&=\delta_{p_2 p_3}\bra{  L_{0 p_4}},\\
		\bra{  L_{p_1 0}}{  A}_{00}&=\bra{  L_{p_1 0}},\\
		\bra{  L_{0 p_2}}{  A}_{00}&=\bra{  L_{0 p_2}},
\end{split}
\end{align}
and if we parametrize the operator as $O=\sum_{ij} \lambda_{ij} E^{ij}_1$, initial conditions read
\begin{equation}
	\bra{L_{p_1 p_2}}R\rangle=\lambda_{p_1 p_2}.
\end{equation}
General solution of bulk, boundary and initial conditions can be represented in three auxiliary spaces $\mathcal{H}_a=\text{End}(\CC^2\otimes \mathcal{F}\otimes \CC^{2q+2})$, where $\mathcal{F}$ is infinitely dimensional. We will make use of four operators from  $\text{End}(\mathcal{F})$
\begin{align}
	\label{auxop}
\begin{split}
	{  a}&=\sum_{k=0}^\infty {  \ket{k}\bra{k+1}},\quad {  a^\dagger}=\sum_{k=0}^\infty {  \ket{k+1}\bra{k}},\quad {  P}_0={  \ket{0}\bra{0}},\\ {  P}_{/0}&=\sum_{k=1}^\infty \ket{  k}\bra{  k},\quad \one=\sum_{k}\ket{  k}\bra{  k}.
\end{split}
\end{align}
Equipped with these definitions, we can write out the solution with the help of matrices $M_{1,2,3}\in \text{End}(\CC^{2q+1})$, 
\begin{align}
	\begin{split}
	M_{p_1 p_2}^{(1)}&=((\ket{0}\bra{0}+\ket{1}\bra{1})\delta_{p_1 p_2}+\ket{1}\bra{2}(1-\delta_{p_1 p_2}))+\ket{p_20}\bra{p_1 0}+\ket{0 p_1}\bra{0p_2}\\
	M_{p_1 p_2}^{(2)}&=(\ket{0}\bra{0}+\lambda_{p_1 p_2}\ket{1}\bra{1})\delta_{p_1 p_2}+\lambda_{p_1 p_2}\ket{1}\bra{2}(1-\delta_{p_1 p_2})+\lambda_{p_10}\ket{p_20}\bra{p_1 0}+\\&+\lambda_{0 p_2}\ket{0p_1}\bra{0p_2},\\
	M_{p_1 p_2}^{(3)}&=\one \delta_{p_1 p_2},
\end{split}
\end{align}
where the basis of the vector space is spanned by 
\begin{equation}
\{\ket{0},\ket{1},\ket{2},\ket{01},\ket{10},\cdots,\ket{0,q-1},\ket{q-1,0}\}.
\end{equation}
Finally, the solution reads
\begin{align}
	\label{sol}
\begin{split}
	{  A}_{00}&={  B}_{00}=\begin{pmatrix}
		{  P}_{/0} M^{(3)}_{00} & 0 \\
		0 & \one M^{(3)}_{00}
	\end{pmatrix},\\
	{ {A}}_{p_1p_2}&=\begin{pmatrix}
		{  P}_{/0}{M}_{p_1 p_2}^{(1)} &  {  P}_{/0}{  a}^\dagger {M}_{p_1 p_2}^{(2)} \\
		0 & {  P}_{/0}{  a}^\dagger{M}_{p_1 p_2}^{(3)}
	\end{pmatrix},\\
	{  B}_{p_1 p_2}&=\begin{pmatrix}
		{  P}_{/0}{  a}{  P}_{/0}{M}_{p_1 p_2}^{(1)} & {  P}_{/0}{M}_{p_1 p_2}^{(2)} \\
		0 & \one{M}_{p_1 p_2}^{(3)}
	\end{pmatrix},
\end{split}
\end{align}
with the boundary vectors
\begin{align}
	\label{bv}
	\begin{split}
		\bra{  L}_{p_1p_2}&=\begin{pmatrix}\bra{{  1}},\lambda_{p_1 p_2}\bra{{  0}}\end{pmatrix}\otimes(\bra{ 1}\delta_{p_1 p_2}+\bra{2}(1-\delta_{p_1 p_2})),\\
		\bra{  L}_{0p_2}&=\begin{pmatrix}\bra{{  1}},\lambda_{0 p_2}\bra{{  0}}\end{pmatrix}\otimes\bra{ {  0}{  p}_2},\\
		\bra{  L}_{p_1 0}&=\begin{pmatrix}\bra{{  1}},\lambda_{p_1 0}\bra{{  0}}\end{pmatrix}\otimes\bra{ {  p}_1{  0}},\\
		\bra{  L}_{00}&=\begin{pmatrix}0,\lambda_{00}\bra{\underline{   1}}\end{pmatrix}\otimes\bra{  0},
\end{split}
\end{align}
\begin{align}
	\ket{  R}=\begin{pmatrix}
		0\\
		\ket{ {0}}
	\end{pmatrix}\otimes \left(\sum_{e} \ket{e}\right).
\end{align}
\section*{References}
\bibliographystyle{iopart-num}
\bibliography{Diffusion_bound}{}
\end{document}